\documentclass[11pt, a4paper]{article}

\pdfoutput=1
\usepackage{jcappub}

\usepackage{graphicx}
\usepackage{datetime}
\usepackage{textpos}
\usepackage{booktabs}
\usepackage{multirow}
\usepackage{color}
\usepackage{lscape}
\usepackage{longtable}

\newcommand{\EC}[1]{{#1}}
\newcommand{\TIM}[1]{{#1}}

\title{Testing the dark matter origin of the WMAP-Planck Haze with radio observations of Spiral Galaxies}
\author{Eric Carlson$^1$, Dan Hooper$^{2,3}$, Tim Linden$^1$, and Stefano Profumo$^{1,4}$}
\affiliation{$^1$ Department of Physics, University of California, Santa Cruz, 1156 High Street, Santa Cruz, CA, 95064, USA}
\affiliation{$^2$ Center for Particle Astrophysics, Fermi National Accelerator Laboratory, Batavia, IL 60510, USA}
\affiliation{$^3$ Department of Astronomy and Astrophysics, University of Chicago, 5640 S. Ellis Ave., Chicago, IL 60637, USA}
\affiliation{$^4$ Santa Cruz Institute for Particle Physics, University of California, Santa Cruz, 1156 High Street, Santa Cruz, CA, 95064, USA}
\emailAdd{erccarls@ucsc.edu}
\emailAdd{dhooper@fnal.gov}
\emailAdd{tlinden@ucsc.edu}
\emailAdd{profumo@ucsc.edu}

\abstract{If the Galactic WMAP radio haze, as recently confirmed by Planck, is produced by dark matter annihilation or decay, similar diffuse radio halos should exist around other galaxies with physical properties comparable to the Milky Way. If instead the haze is due to an astrophysical mechanism peculiar to the Milky Way or to a transient event, a similar halo need not exist around all Milky Way ``twins''. We use radio observations of 66 spiral galaxies to test the dark matter origin of the haze. We select galaxies based on morphological type and maximal rotational velocity, and obtain their luminosities from a 1.49 GHz catalog and additional radio observations at other frequencies. We find many instances of galaxies with radio emission that is less than 5\% as bright as naively expected from dark matter models that could produce the Milky Way haze, and at least 3 galaxies that are less than 1\% as bright as expected, assuming dark matter distributions, magnetic fields, and cosmic ray propagation parameters equal to those of the Milky Way. For reasonable ranges for the variation of these parameters, we estimate the fraction of galaxies that should be expected to be significantly less bright in radio, and argue that this is marginally compatible with the observed distribution. While our findings therefore cannot rule out a dark matter origin for the radio haze at this time, we find numerous examples (including the Andromeda Galaxy) where, if dark matter is indeed the origin of the Milky Way haze, some mechanism must be in place to suppress the corresponding haze of the external galaxy. We point out that Planck data will offer opportunities to improve this type of constraint in a highly relevant frequency range and for a potentially larger set of candidate galaxies.}

\begin{document}
\maketitle

\section{Introduction}
\label{sec:introduction}

Data from NASA's {\em Wilkinson Microwave Anistropy Probe} (WMAP) have fostered great advances in both precision cosmology and in the understanding of processes associated with the interstellar medium of our own Galaxy. WMAP also brought about unexpected observational results, including the detection of spinning dust emission \cite{draine_lazarian_a, draine_lazarian_b, deoliveiraetal, finketal} and a mysterious diffuse microwave emission extending 20$^{\circ}$ around the Galactic Center, known as the ``WMAP haze'' \cite{fink04}. While originally associated with free-free emission \cite{fink04}, follow-up analyses showed that the haze spectrum was too soft to be attributed to free-free emission, but too hard to match the diffuse Galactic synchrotron emission produced by relativistic cosmic ray electrons \cite{DF08}.

While some authors have criticized the original analysis procedure \cite{mertsch_sarkar}, or the synchrotron nature of the haze \cite{gold09}, both arguments have been rebutted in detail \cite{pietrobon, dobler12}. Excitingly, the {\em Planck} mission has recently presented conclusive evidence in favor of the existence of the microwave haze \cite{planck}, employing superior background rejection methods to those available to the WMAP team. Specifically, the {\em Planck} data analysis proceeded along two distinct component separation techniques, one following the original WMAP template analysis of Ref.~\citep{DF08}, and a more sophisticated Bayesian approach based on Gibbs sampling. Both techniques confirmed the existence of an anomalous diffuse haze in the Galactic Center (GC) region across multiple frequencies, with a spectral index $\beta_H=-2.55\pm0.05$, consistent with that found from WMAP data. Furthermore, future {\em Planck} studies will analyze the polarization of the haze, providing an even more stringent test of the synchrotron nature of this emission.

Several authors have considered dark matter annihilation as a possible source of the Galactic haze (see e.g. Ref.~\cite{fink_dm_04, hooper07, lin10, dobler11, lindenhooper, lindenprofumo,add13,add9,add5,add1,add7,add3,add2}) \EC{as well as in local extragalactic dark matter searches \cite{add10,add11,add6,add4,add8,add12}}. The dark matter interpretation is consistent with theoretically well motivated ranges for the mass and annihilation cross section of weakly interacting massive particles, common to many extensions of the standard model of particle physics \citep{dmreview, dmreview2, uedreview}. Furthermore, it was noted recently that the WMAP haze is consistent with the annihilations of light ($m\sim10$ GeV) dark matter candidates which are also capable of accounting for a number of other reported observations~\citep{lindenhooper}, including direct detection signals from DAMA/LIBRA~\citep{dama}, CoGeNT~\citep{cogent1, cogent2} and CRESST-II~\citep{cresst2}, as well as the spatially extended spherically symmetric excess of $\gamma$-rays observed around the GC by the Fermi-LAT~\citep{hooper_goodenough, hooper_linden_gc, abazajian_kaplinghat}, the extremely hard synchrotron spectrum observed in multiple filamentary arcs surrounding the GC~\citep{filamentary_arcs}, and the isotropic radio excess observed by the ARCADE collaboration~\citep{arcade} (see Ref.~\citep{fornengo_2011, dan_empirical_case} for further discussion of these observations and experiments).

Recent observations indicate that the observed morphology of the microwave excess haze exhibits a strong correlation with a diffuse feature discovered at GeV energies by the {\em Fermi} Large Area Telescope, known as the Fermi bubbles \cite{suetal} (previously known as ``Fermi haze'' \cite{fermihaze}, see also \cite{timandstefano}). This suggests a common origin for the observed emissions at radio and gamma-ray energies~\citep{dobler_radio_gamma_connection}. Hypotheses as to the origin of this additional cosmic ray population include anomalous enhancements to the supernova activity in the relevant region and time  \cite{biermannetal}, peculiar cosmic ray propagation in the region of interest, perhaps associated with Galactic winds \cite{crockeraharonian}, central nuclear activity fueling cosmic ray jets \cite{guoetal}, or an additional population of cosmic ray electrons distinct from the ordinary diffuse cosmic ray population in the Galactic plane~\citep{lin10}. \EC{While high mass dark matter models are capable of simultaneously explaining energy \emph{spectrum} of the WMAP haze and Fermi Bubbles, no dark matter model is capable of accommodating the sharp edges observed at high galactic latitude~\citep{dobler11}, suggesting a much more complicated story if dark matter is to remain a primary progenitor of the WMAP haze.}

In this \emph{paper}, we point out that if the WMAP haze results from dark matter annihilation, a radio signal with a similar luminosity and morphology should be emitted from other galaxies, so long as these galaxies are comparable to our own in terms of size and other physical characteristics. This comparison may not hold in scenarios where the haze is of astrophysical origin (e.g. in the case that the haze is powered by episodic nuclear activity in the GC). The purpose of this study is to identify a suitable set of galaxies similar to the Milky Way, estimate the radio ``haze'' expected from these candidate galaxies assuming a dark matter annihilation origin and benchmark values for the parameters governing the propagation of charged leptons in each galaxy, and then employ radio observations to obtain constraints on the dark matter annihilation origin of the Milky Way haze.

In order to interpret the results of this comparison, it is necessary to estimate the uncertainties in the theoretical expectation for the dark matter radio haze in external galaxies. Relatively little is known about propagation of cosmic rays, magnetic fields, or dark matter distribution in external galaxies. Any or all of these characteristics could vary considerably from galaxy-to-galaxy, even among those galaxies that are similar in size and morphology to the Milky Way. We therefore study in detail the expected range of variation in the radio luminosity of the dark matter radio haze originating from such variations.

Our study is structured as follows. In the next section, we describe the selection criteria we employ to select the relevant set of galaxies to investigate. In Sec.~\ref{sec:sysUncertanties}, we address the systematic uncertainties on the estimate of the haze luminosity. In Sec.~\ref{sec:theoretical_uncertainties} we produce a Monte Carlo model indicating the expected variation in the dark matter induced synchrotron signal due to these systematic uncertainties. In Sec.~\ref{sec:fluxcomparison}, we compare our predictions with radio data. Lastly, we discuss implications of our results and future directions in Sec.~\ref{sec:conclusions}.

\section{Galaxy Sample Selection}
\label{sec:Selection}
In order to identify galaxies with size and morphology similar to the Milky Way, we utilize the 1.49 GHz radio data from the Condon Atlas of Spiral Galaxies (hereafter refered to as CA)~\citep{Condon:1987}. We obtain morphological information for each galaxy by cross-referencing the CA with the HyperLeda database\footnote{The HyperLeda database may be found at \href{http://leda.univ-lyon1.fr/}{http://leda.univ-lyon1.fr/}}, and restrict our analysis to galaxies with a morphological type  $1.5<t<4.0$ \footnote{Mapping between Hubble classification and de Vaucouleurs morphological type index may be found at \href{http://leda.univ-lyon1.fr/leda/param/t.html}{http://leda.univ-lyon1.fr/leda/param/t.html}}, which limits the uncertainties associated with edge-on galaxies, for which morphological properties can be difficult to determine. We note that this conservative cut loosely includes type Sab-Sbc galaxies, \EC{where the Milky Way is typically classified near Sbc \cite{MWTYPE}. }

Apart from well-studied galaxies such as Andromeda, no specific information on the dark matter density profile is available for the galaxies in the CA. Most of these galaxies, however, have estimates for the values of the maximum rotational velocity, given by observations of 21 cm line widths. These values are highly correlated with the total enclosed mass in each galaxy. We place a cut on the maximum rotational velocity of 180 km/s$<v_{rot}<280$ km/s to ensure a mass and size compatible with the Milky Way\EC{, where $v_{rot}\approx 235$ \cite{MWROT}. }

\EC{After each of the above cuts is applied, the resulting sample contains 66 galaxies which are used throughout this analysis.} We then rank the galaxy candidates based on the observed radio flux in two ways. First, we calculate the total radio luminosity at 1.49~GHz given the total observed radio flux ($S$) reported in the CA and the best-fit distance estimate as given by the NASA/IPAC Extragalactic Database (NED)\footnote{The NASA/IPAC Extragalactic Database (NED) is operated by the Jet Propulsion Laboratory, California Institute of Technology, under contract with the National Aeronautics and Space Administration.}. Second, we calculate the luminosity 2~kpc above or below the Galactic plane by using the peak flux measurement ($S_P$) reported by the CA, as well as the elliptical Gaussian defined by the FWHM at 1.49 Ghz along the major and minor axes, with the major axis assumed to lie along the Galactic plane as calculated by optical data. \EC{The resulting list is then sorted in ascending order of their relative luminosity, found by multiplying the flux times the square of the distance, and the two rankings are averaged. In Table~\ref{tab:selected} in the appendix we list each of the 66 galaxies meeting our selection criteria along with their luminosity ranking.  In Table~\ref{tab:select}, we list the seven \emph{lowest luminosity} galaxies which we will analyze in more individual detail than the full 66 galaxy sample.  Ultimately, this subset of the 7 galaxies with the lowest cosmic ray background will provide the most stringent limits on a dark matter induced haze.} \TIM{We note that all statistical comparisons showing the percentage of galaxies which are underluminous are calculated using the full population of 66 galaxies, to remain consistent with the total distribution from which galaxies are drawn. We also note that all of our ensuing statistical statements are made by employing the large, 66 galaxy sample, but our constraints are derived from the most promising 7-galaxies sub-sample. Since we would have gotten even more stringent constraints had we considered more than the 7 most promising objects, we deem our results as conservative (i.e. they in principle can be improved upon).}

\begin{table}[t]
\begin{center}
\scriptsize
\begin{tabular}{cccccccccccccccccc}
\toprule
Name                   &       Type        &         $t$         &       $i$        &       $v_{\rm rot}$        & $S$ & $S_p$  & D  & Rank$_L$ & Rank$_{2{\rm kpc}}$ & {Mean Rank} \\
	                   &       	           &                     &       $[\rm deg.]$  &       [km/s]        & [mJy] & [mJy]  & [Mpc] &   &  &\\ \hline
NGC 4448                                 &  SBab             &  1.8$\pm 0.6$              &  69.00~\cite{Heald:2011} &  221.54           & 1 & 0.9        	 & 13.1  & 1 & 2 & 1.5 \\ 
NGC 4698                                 &  Sab              &  1.7$\pm 1.0$              &  73.44            &  201.13           & 0.6  			 &     				     & 21.9  & 2 & & 2 \\ 
NGC 4394                                 &  SBb              &  3 $\pm 0.4$               &  16.55            &  255.13           & 0.7 & 0.6  & 21.9 &  3 & 3 & 3 \\ 
NGC 7814                                 &  Sab              &  2$\pm 0.2$                &  90.00~\cite{Fraternali:2011} &  230.9            & 1.1 & 1.1         & 25.8  & 4 & 6 & 5 \\ 
NGC 1350                                 &  Sab              &  1.9$\pm 0.6$       &  64.79            &  199.87           & 1.1 & 0.8         & 30.2  & 5 & 7 & 6 \\
NGC 0224                                 &  Sb               &  3$\pm 0.4$                &  72.17            &  256.7            & 8400 & 14.5         & 0.7 & 12 & 1 & 6.5 \\ 
NGC 2683                                 &  Sb               &  3$\pm 0.3$                &  82.79            &  202.92           & 65.9 & 14.6  & 8 &  13 & 5 & 9 \\ 
\bottomrule
\end{tabular}
\end{center}
\caption{Candidate galaxies from the Condon Atlas.  $t$, $i$ , $v_{\rm rot}$ , $S$, $S_p$, refer to the de Vaucouleurs morphological type, inclination, maximum rotational velocity, total flux, and peak flux, respectively.  Rank$_L$ (Rank$_{2{\rm kpc}}$) ranks the total (respectively, 2 kpc off disk) luminosity, in ascending order.  Unless otherwise noted, inclination, type, and rotational velocity are from the HyperLeda database. See Table~\ref{tab:selected} for complete listing of all 66 selected galaxies.}
\label{tab:select}
\end{table}

In order to estimate the physical size of the cosmic ray diffusion region, we use as a proxy the projected size of the major axis of each galaxy, based on the distances obtained from NED and an angular size from HyperLeda, defined as the isophotal level 25 mag/arcsec$^2$ in the B-band \cite{patureletal1991}. 
In Table~\ref{tab:sizes}, we list the mean distance for the galaxy sample, as given in the NED database. For each galaxy in the sample, we extrapolate the physical size of the Galaxy's major axes. We also report the ratio of the major-to-minor axis in the last column, although we assume cylindrical symmetry for the simulated diffusion region.

\begin{table}[t]
  \scriptsize
  \centering
  \begin{tabular}{cccccccc}
    \toprule
    Name & NED Distance & Major Axis & Axes Ratio \\
     & [Mpc] & [kpc] &  \\\midrule
  NGC 0224         & 0.7 & 60.2 & 2.55 $\pm$ 0.29 \\ 
  NGC 1350         & 20.9 & 52.4 & 1.99 $\pm$ 0.15 \\
  NGC 2683         & 10.2 & 46.9 & 3.51 $\pm$ 0.42 \\
  NGC 4394         & 16.8 & 28.4 & 1.04 $\pm$ 0.07 \\ 
  NGC 4448         & 13.0 & 10.0 & 1.53 $\pm$ 0.15 \\ 
  NGC 4698         & 23.7 & 43.8 & 2.45 $\pm$ 0.16\\ 
  NGC 7814         & 17.2 & 36.4 & 2.32 $\pm$ 0.14 \\ 
    \bottomrule
  \end{tabular}
  \caption{Distances and sizes of the candidate sample galaxies.}
  \label{tab:sizes}
\end{table}

We conclude that all of our candidate galaxies have comparable diffusion region sizes, with the possible exceptions of NGC 4448 and NGC 4394, which exhibit a potentially smaller radius and a smaller major-to-minor axis ratio.  These galaxies also have the largest distance uncertainties, directly affecting the reported semi-major axis.

\section{The Dark Matter Haze in External Spiral Galaxies: Systematic Uncertainties}
\label{sec:sysUncertanties}

Although we employ morphological information to choose a sample of galaxies similar to the Milky Way, there are several other parameters capable of impacting the radio luminosity of a dark matter haze which can vary significantly from galaxy-to-galaxy. Generically, these parameters break down into three categories: 
\begin{enumerate}
\item parameters that control the dark matter density profile, 
\item parameters that control the diffusion of cosmic rays, and 
\item parameters that control the magnetic field strength in each sample galaxy. 
\end{enumerate}
In Table~\ref{tab:paramrange} we show the default, maximum and minimum value for each parameter used throughout this work, and in Figure~\ref{fig:scanHist} we show the effect of changes in each individual parameter on the synchrotron luminosity due to dark matter annihilation at 1.49~GHz. In the following, we discuss the variations in the radio haze produced by of these parameters in detail. 

\subsection{Dark Matter Density Profile}
\label{subsec:density_profile}

As the dark matter annihilation rate is proportional to the square of the dark matter density, the total synchrotron luminosity depends not only on the total dark matter mas  (which can be reasonably estimated from galactic rotation measurements) but also on the density distribution of dark matter in each galaxy. In this work, we assume a generalized Navarro-Frenk-White (NFW) profile~\citep{nfw} with a radial density profile given by:

\begin{equation}
\rho(r) = \rho_s(\frac{r_s}{r})^\alpha(1+\frac{r}{r_s})^{-3+\alpha}
\end{equation}

\noindent where $r_s$ is a scale radius which governs the turnover to the $r^{-\alpha}$ profile, $\rho_s$ is a dark matter density normalization constant, and $\alpha$ governs the inner slope of the dark matter density. In the original NFW profile, $\alpha$ = 1, in agreement with the results of certain dark matter-only simulations of structure formation, e.g. Ref.~\citep{aquarius}. However, recent simulations which include the effects of baryons on galaxy evolution have found that the cooling and contraction of the baryonic density profile tend to steepen the dark matter density profile in the central regions~\citep{adiabatic_contraction}. This effect, known as baryonic contraction, generally leads to values of $\alpha$ in a range between 1.2-1.7, depending on the degree of contraction. In contrast, other numerical simulations which include a large degree of baryonic feedback have found that baryons can flatten the inner slopes of the dark matter profile, producing distributions with a flat density core as large as 1~kpc in size~\citep{governato_dark_matter_cores}. In our simulations we allow for r$_s$ to change by a factor of 2 from our default choice of 22.0~kpc, and we allow $\alpha$ to change by a factor of 1.5 from a default value of 1.0. We normalize the dark matter density profile the dark matter density at the solar position (R$_\odot$ = 8.5~kpc) to a range within a factor of 5 of the central value of 0.3~GeV~cm$^{-3}$~\citep{bovy_tremaine}. 

\begin{figure}[t]
\begin{center}
	\includegraphics[width=\textwidth]{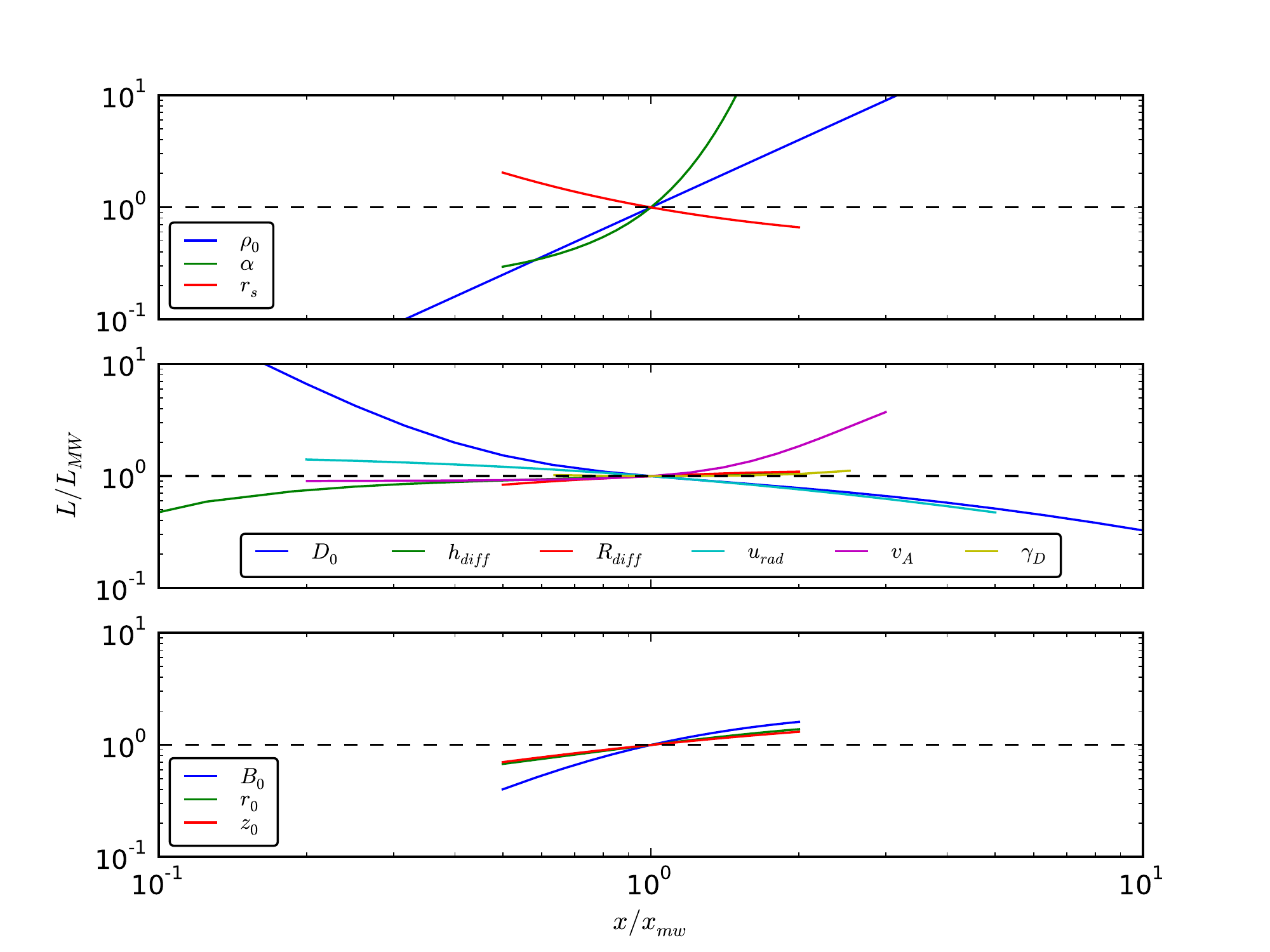}
\end{center}
\caption{The effect of variations in the nuisance parameters on the luminosity of the dark matter synchrotron haze at 1.49 GHz in units of the benchmark value, $L_{\rm MW}$.  We note that variations in $h_{diff}$ closely floow  See Table~\ref{tab:paramrange} and text for more details.}
\label{fig:param}
\end{figure}

Figure~\ref{fig:param} (top) illustrates that the haze luminosity depends strongly on the assumed value for $\alpha$, increasing precipitously for profiles with $\alpha>1$. We note that this will affect our sample producing a logarithmically asymmetric distribution in luminosity and favoring large radio fluxes. We also note only a marginal dependence on the value of $r_{s}$ as the annihilation which produces the radio haze is largely confined to within the inner several kpc around the center of each galaxy.

\begin{table}[t]
  \scriptsize
  \centering
  \begin{tabular}{cccccc}
    \toprule
   Parameter & [units] & $c_{\rm f.f.}$ & Central & Min & Max  \\\midrule
$B_0$ & $\mu$G & 2 & 60 & 30 & 120 \\
$r_0$ & kpc & 2 & 4.0 & 2.0 & 8.0 \\
$z_0$ & kpc & 2 & 1.8 & 0.9 & 3.6 \\\midrule
$D_0$ & cm$^2$s${^{-1}}$ & 10 & 1.0~x~10$^{29}$ & 1.0~x~10$^{28}$ & 1.0~x~10$^{30}$ \\
$h_{\rm diff}$ & kpc & 10-2 & 16 & 1.6 & 32 \\
$R_{\rm diff}$ & kpc & 2 & 20 & 10 & 40 \\
$u_{\rm rad}/u_{\rm rad,MW}$ &  & 5 & 1 & .2 & 5 \\
$v_{A}$ & km s$^{-1}$ & 1 & 25 &25 & 25\\ 
$\gamma_{D}$ &  & 1 & .33 & .33 & .33\\ \midrule
$\rho_0$ & GeV cm$^{-3}$ & 5 & 0.30 & 0.06 & 1.50 \\
$r_s$ & kpc & 2 & 22 & 11 & 44 \\
$\alpha$ & & 1.5 & 1.0 & 0.67 & 1.5 \\
    \bottomrule
  \end{tabular}
  \caption{Nuisance parameters employed to estimate the range for a dark matter induced haze in external spiral galaxies. Note that we also place an additional constraint on each simulation that the diffusion height must not exceed the simulated diffusion radius.  \EC{The Alfven velocity $v_A$ and the energy dependence index, $\gamma_D$, of the diffusion coefficient are held fixed at the default values for all analyses, though the effects on the synchrotron luminosity are presented in Figure \ref{fig:param}.}}
  \label{tab:paramrange}
\end{table}

\subsection{Cosmic Ray Diffusion}
As charged particles propagate through the galaxy, they are deflected by the turbulent magnetic field structure and proceed in a random walk that can be described as a standard diffusive process. Since Galactic magnetic fields are thought to be produced primarily by magnetohydrodynamic turbulence, they are highly dependent on astrophysical parameters such as the distribution of molecular clouds, ionized gas, star formation rates, etc. which vary wildly between galaxies~\citep{galactic_magnetic_fields}. Additionally, the nature of charged particle propagation depends sensitively on the scale of the magnetogydronamic fluctuations compared to the gyroradius of the charged particle in the magnetic field.  This is a local phenomena which is not uniform across different regions of the Galaxy.

Although an exact solution to the diffusion equation for a galaxy is not attainable, cosmic ray propagation codes exist that allow for realistic estimates of charged particle diffusion in the Milky Way. The predictions of these codes can be tested by observable quantities such as the ratios of radioactive secondaries to primary species observed at the solar position, large scale gamma-ray emission, and other diffuse Galactic radiation backgrounds. The \emph{Galprop} code stands at the forefront of this effort, performing numerical calculations of particle propagation and of the resulting multi-wavelength emission, taking into account effects such as the diffusion, reacceleration, and convection of cosmic rays, the intensity as well as the morphology of the Galactic magnetic and interstellar radiation fields, and the morphology of molecular gas~\cite{galprop}. In this analysis, we employ \emph{Galprop} (v. 54) to study the effect of varying four parameters, which previous analyses of the WMAP haze in the Milky Way determined to have the largest effect on the synchrotron signal due to dark matter annihilation~\citep{lindenprofumo}. Specifically we examine changes in the mean diffusion constant, $D_0$, whose benchmark value we take to be 1.0~$\times$~10$^{29}$~cm$^2$~s$^{-1}$, the radius ($R_{\rm diff}$) and half-height ($h_{\rm diff}$) of the diffusion region, which we take to be 20~kpc and 16~kpc for our central values, and the normalization of the inter-stellar radiation field, which we take to be unity in units of the default \emph{Galprop} model.

Figure~\ref{fig:param} (middle) examines the changes in the mean Milky Way luminosity for variations in our diffusion setup, finding only an minimal effect from changes in the size of the diffusion region, on the order of 10\% for regions double the default size. However, changes in the diffusion constant produce a larger effect, especially in the regime of low diffusion constants where electrons created by dark matter annihilations are efficiently trapped in the galaxy. Large increases in the strength of the interstellar radiation field also greatly decrease the luminosity of the synchrotron haze, as charged leptons lose a significant fraction of their power to inverse Compton scattering as opposed to synchrotron.  \EC{Variations of the default the Alfven velocity, $v_A$=25 km/s, show the effect the electron re-acceleration which negligibly decreases the luminosity at lower values and increases as it is raised to 100 km/s. The energy index of the diffusion coefficient, $\gamma_D$, is also shown and indicates that the default value of $\gamma_D = 0.33$ produces the minimum luminosity and variations are positive over the rest of the physical domain ($.2\leq \gamma_D \leq 7$).  Due to the inherent difficulty in constraining these parameters physically, both the Alfven velocity and the energy index of diffusion are held fixed at their conservative default choices for the remainder of this paper.}

\subsection{Galactic Magnetic Fields}
In addition to influencing the propagation of charged cosmic rays, Galactic magnetic fields also control the intensity and spectrum of synchrotron radiation produced by cosmic ray electrons as they travel through the interstellar medium. Unfortunately, the mean intensity of magnetic fields is highly uncertain, even in the case of the Milky Way. However, constraints from the rotation measures of pulsars place the mean magnetic field strength in the Milky Way to be approximately 4-6~$\mu$G~\citep{milky_way_magnetic_field}. In this paper, we adopt the a magnetic field model described by the functional form $B(r,z)=B_0 \exp(-r/r_0)\exp(-z/z_0)$, and vary the parameters of this expression as described in Table~3.

In Figure~\ref{fig:param} (bottom) we show the variation in the dark matter synchrotron luminosity for changes in the strength of the magnetic field in a given galaxy. We find only moderate changes when we shift the overall amplitude of the magnetic field, which is primarily due to the large fraction of electron energy which is lost to synchrotron for all magnetic field strengths. The dependence of the total synchrotron intensity on the radial and zenith scale of the magnetic field is even more negligible, due to the fact that the synchrotron intensity is dominated by the very inner region around each galaxies' center.

\section{Monte Carlo Modeling of Systematic Uncertainties}
\label{sec:theoretical_uncertainties}
\EC{In order to estimate a probability distribution for the luminosity of each external galaxy, we produce a Monte Carlo sample, varying each parameter listed above within the minimum and maximum values provided in Table 3. We sample all of parameter space linearly except for the diffusion constant $D_0$ and the interstellar radiation density $u_{rad}$ which are sampled logarithmically. In the case of the diffusion constant the effect of logarithmic sampling can bias the distribution toward higher luminosities as a low diffusion constant more efficiently traps  electrons and positrons in the galaxy.  While this does produce somewhat less conservative constraints,  linear sampling would have produced unrealistically large diffusion constants on average and we find that the logarithmic results are generically more compatible with the range of values preferred by cosmic-ray data.  The remaining parameters are either sub-dominant drivers of the luminosity, or are only varied by a factor of 5 or less.  In these cases, we do not expect Bayesian priors to have a significant effect.}  We provide an additional constraint on the parameters for the dark matter density profile, specifying that the total mass enclosed within 100~kpc fall between 0.25 and 2.0 times the value for the default Milky Way Halo. This is consistent with the fact that we select galaxies in our sample within a relativity narrow mass range as fixed by our constraint on v$_{\rm rot}$.

In order to estimate a probability distribution for the luminosity of each external galaxy, we produce a Monte Carlo sample, varying each parameter listed above within the minimum and maximum values provided in Table 3. We sample all of parameter space linearly except for the diffusion constant $D_0$ and the interstellar radiation density $u_{rad}$ which are sampled logarithmically. In the case of the diffusion constant the effect of logarithmic sampling can bias the distribution toward higher luminosities as a low diffusion constant more efficiently traps  electrons and positrons in the galaxy.  While this does produce somewhat less conservative constraints,  linear sampling would have produced unrealistically large diffusion constants on average and we find that the logarithmic results are generically more compatible with the range of values preferred by cosmic-ray data.  The remaining parameters are either sub-dominant drivers of the luminosity, or are only varied by a factor of 5 or less.  In these cases, we do not expect Bayesian priors to have a significant effect.

Figure~\ref{fig:scanHist} shows the distribution in the total luminosity calculated at 1.49~GHz for our Monte Carlo sample, illustrating  the luminosity due both to dark matter annihilation as well as the cosmic ray luminosity for a galaxy with a cosmic ray injection rate equal to that of the Milky Way. A time-independent change in the cosmic ray injection rate would simply linearly shift the cosmic ray induced radio luminosity of a given galaxy. We also show for comparison the distribution of luminosities for all 66 ``Milky-Way-like" candidates taken from the CA catalog, using the best-fitted distance given by NED. In Figure~\ref{fig:scanHistContour}, we show the distribution of dark matter-induced luminosities as determined by our Monte Carlo as evaluated along the Galactic plane at a distance of $r=5$ kpc from the GC (red line), and above the Galactic plane at a height $z=1$ kpc (blue line).  Also repeated for comparison is the integrated luminosity due to dark matter.  

At this point, we note several key results:

\begin{figure}[tbp]
\begin{center}
	\includegraphics[width=\textwidth]{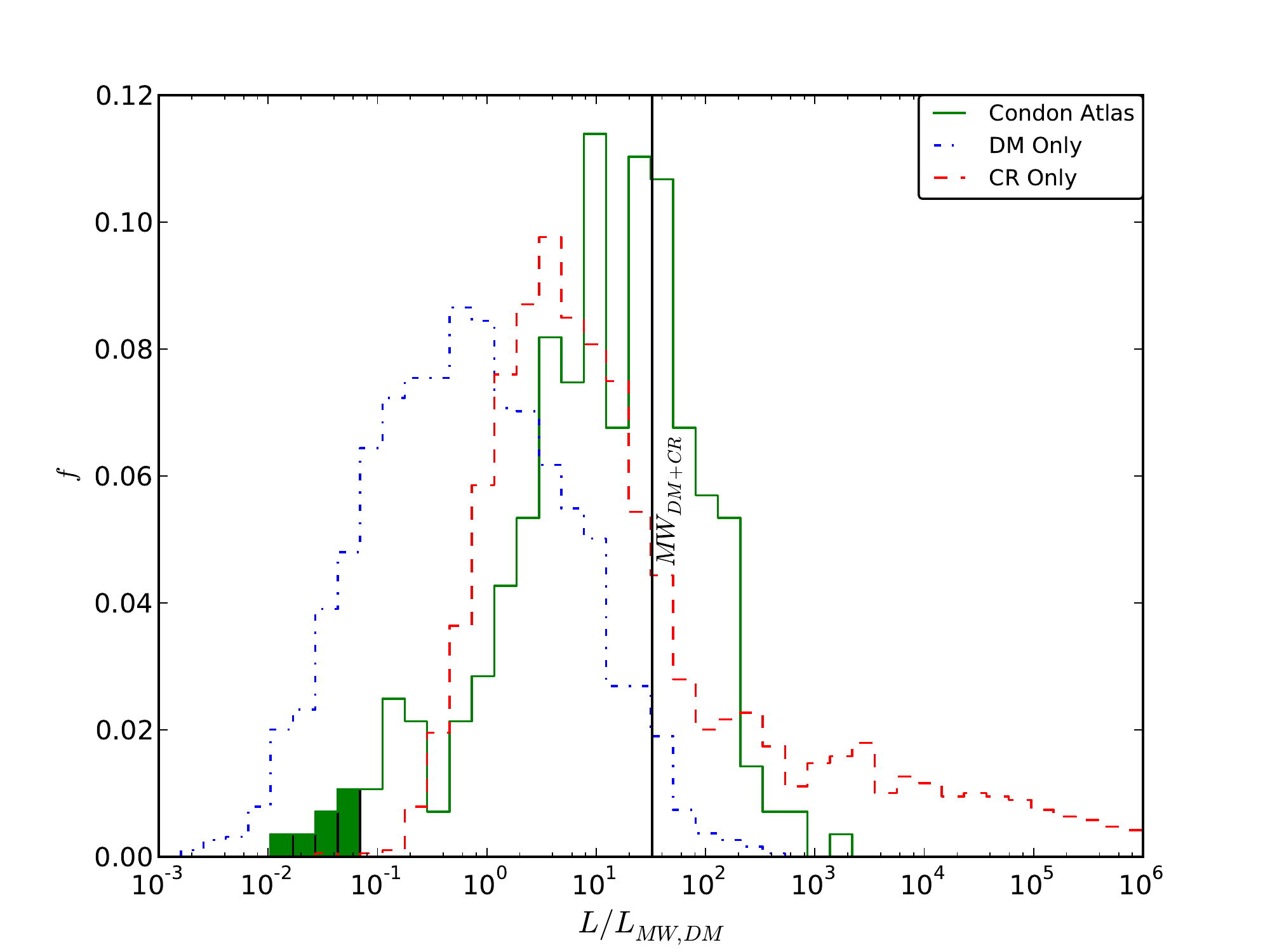}
\end{center}
\caption{Distribution of 1.49 GHz luminosity ratio $L/L_{MW,DM}$ where $L$ is the total integrated luminosity and $L_{MW,DM}$ is the luminosity due to dark matter for the canonical Milky Way model.  The benchmark luminosity for dark matter plus cosmic rays is indicated by the vertical line.  The simulated sample contains 2000 runs randomly distributed in the parameter space of tab.~\ref{tab:paramrange} with mass restricted to lie within $0.25-2$ times the mass of the Milky Way. We plot the contributions due to dark matter only (dot-dashed blue) and to cosmic rays only (dashed red) against extrapolated luminosities of 66 Condon Atlas galaxies meeting morphological cuts described in Section~\ref{sec:Selection} (solid green).  Only $11.2\%$ of the simulated dark matter-induced haze luminosities are a factor 20 smaller than our benchmark value, and only $1.42\%$ are a factor 100 smaller.  \EC{Shown in shaded green are the 7 lowest background galaxies selected from the Condon Atlas.}}
\label{fig:scanHist}
\end{figure}

\begin{figure}[tbp]
\begin{center}
	\includegraphics[width=\textwidth]{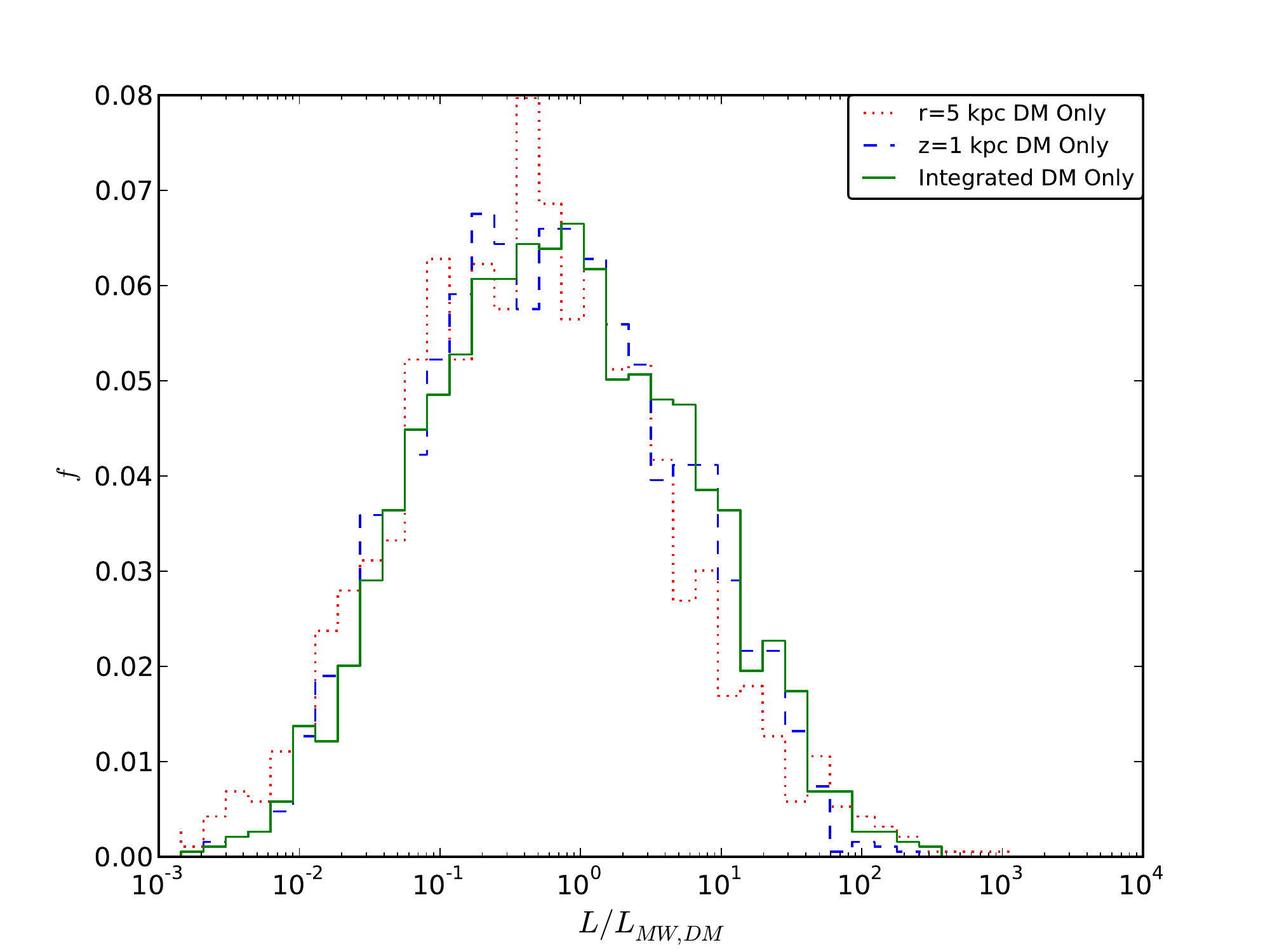}
\end{center}
\caption{Distribution of 1.49 GHz luminosity ratio $L/L_{MW,DM}$ where $L$ is the luminosity evaluated at a radius $r=5$ kpc along the Galactic plane (dotted red), height $z=1$ kpc above the Galactic Center (dashed blue) and $L_{MW,DM}$ is the luminosity due to dark matter for the canonical Milky Way model.  The sample is identical to that of Figure \ref{fig:scanHist} and we have repeated the integrated dark matter component (solid green).}
\label{fig:scanHistContour}
\end{figure}

\begin{enumerate}
\item The total synchrotron luminosity from dark matter is almost invariably found to be subdominant to that from astrophysical cosmic ray sources, typically by more than one order of magnitude. We note from Figure~\ref{fig:param} that this conclusion depends primarily on the dark matter profile, as the cosmic ray injection intensity is not altered in our simulations. This also implies that setting constraints on dark matter annihilation in external galaxies by assuming that the radio luminosity due to astrophysical sources is negligible is an extremely conservative approach, as the radio luminosities of most galaxies is expected to be dominated by cosmic rays from astrophysical sources.

\item Our modeled astrophysical synchrotron radiation has a more pronounced high luminosity tail than is seen in the distribution of the CA galaxies, implying either that the range in the parameter space for cosmic ray diffusion is tilted towards very radio bright systems, or that the cosmic ray injection sources in our model galaxies are typically smaller than in the Milky Way. This is not true for our distribution of dark-matter induced emission, however, which is symmetrically distributed in log-space around the luminosity of the Milky Way. 

\item Only 11.2\% of the simulated integrated dark matter-induced haze luminosities are a factor of 20 smaller than our benchmark value, and only 1.4\% are a factor of 100 smaller. This implies that the dark matter haze is a reasonably resilient feature among a large array of dark matter density profiles and diffusion scenarios. 

\item The simulated dark matter hazes can also be parameterized by the luminosity evaluated at 5 kpc along the Galactic plane and 1 kpc above the Galactic plane.  In the r (z) directions, only 15.2\% (12.3\%) of simulations are suppressed by a factor 20 and 3.8\% (1.7\%) are suppressed by a factor 100.  In the next section we find that the strongest constraints are typically set by comparing the flux along the major-axis.  For high inclination sources this is coincident with the radial direction considered here. 
\end{enumerate}

\section{Constraints from Radio Observations}\label{sec:fluxcomparison} 

\subsection{Methodology}
In addition to the above estimates of the total synchrotron power for a wide sample of values for the nuisance parameters, the \emph{Galprop} code predicts in detail the morphology of the synchrotron emission stemming from dark matter annihilation. \EC{For each of the 7 lowest cosmic-ray background CA candidate galaxies listed in Table \ref{tab:select} this morphological information allows us to set constraints in two additional ways besides considering only the total integrated luminosity: }
 
\begin{itemize}
\item First, we can set constraints by comparing the peak flux contour in each galaxy against the predicted dark matter flux at the Galactic Center. 
\item Second, we can compare the flux above or below the Galactic plane, where the dark matter produced haze is expected to be most significant compared to the synchrotron flux from cosmic rays. In addition we can compare the flux along the galactic plane in each direction.  In order to do this, we calculate the flux at the outermost observed radio contour for each galaxy and compare it to the modeled dark matter flux. We note that Figure~\ref{fig:scanHistContour} shows that the scale of the luminosity distributions from our dark matter models is expected to be similar in all cases.
\end{itemize}

\noindent For the simulations presented in this section, we set all relevant parameters to the central values quoted in Table~\ref{tab:paramrange}, except for the radial size of the diffusion region, which we set to the size of the major axis as given in Table~\ref{tab:sizes}.  We then use {\em Galprop} to calculate the synchrotron emission throughout our model galaxy as a function of the radius and height, and integrate the emission over the line-of-sight based on the observed inclination of each candidate galaxy, correcting for the instrument beam-width for each galaxy.  We note two subtle points in this analysis.  

Firstly, the diffusion height employed in this analysis is extremely large. Although this will produce a slightly overluminous synchrotron signal, the results of Section \ref{sec:sysUncertanties} show that this is a small effect until the diffusion height becomes less than 2 kpc.  This follows from the exponential falloff of the magnetic field with a default characteristic height of 1.8kpc, which is effectively halved when considering the synchrotron luminosity ($L_{sync} \propto B^2$).  We have confirmed in detail that the effect of changing the diffusion height has only an extremely suppressed effect until $h_{diff}\lesssim2$ kpc.  Very small diffusion heights can have a more pronounced effect when evaluating the off-peak flux along the radial direction because the electrons and positrons can escape through the top and bottom of the diffusion region before they are able to diffuse to large radii.  For flux measurements at $r=5$ kpc along the radial direction, the flux can vary up to $\sim 20\%$ when the diffusion height becomes less than 4 kpc.  However, this is a relatively small effect occurring under special circumstances and is not expected to play a significant role in our results.  

The second point is to consider the possibility of correlations between the dark matter and cosmic-ray luminosity components, i.e. does a galaxy with a low cosmic-ray signal necessarily have a low dark matter signal?  To answer this we can infer from Figure \ref{fig:param} that the dark matter halo parameters produce the largest changes to the DM induced haze luminosity, with the exception of the diffusion constant $D_{0}$.  We therefore expect that the DM and CR haze luminosities should be effectively decoupled, as the dark matter component depends primarily on the dark matter density parameters.  Detailed checks have shown that this is indeed the case, and that these two luminosities are uncorrelated at first order\footnote{We thank the Referee for bringing this cross-check to our attention.}.

In order to compare the resulting flux contour map to observations of each CA candidate galaxy, we take into account the distance to each candidate galaxy and the related observational uncertainties. Here, we only consider the central value for the distance measurement when the flux error due to the distance uncertainty is less than 25\% of the standard deviation due to the systematic uncertainties discussed in Section~\ref{sec:sysUncertanties}. This condition does not hold, however, for the galaxies NGC 2683, NGC 4698 and NGC 4448.  For these three galaxies, we provide flux contour-maps for the average, as well as the $\pm 1 \sigma$ distance measurements. We note that in the case of NGC 4448, only three distance measurements exist in the literature: we thus show results for the minimum, average, and maximum distance measurements reported in NED. For each model galaxy, we match our minimum contour to the minimum contour which is observationally reported, and then plot contours in flux increments of 2$^{n/2}$ for integers $n$, as given by the CA catalog. For most galaxies, the lowest contour corresponds to $n=-3$, since the minimum observable flux/beam is relatively distance independent for resolved galaxies. However, the fluxes for Andromeda are reported with fluxes starting at $n=-2$. 

In order to calculate the peak flux contour from our \emph{Galprop} models, we must employ a minimum angular region in which to calculate the peak flux, and this region must exceed the size of the contour region resolved by the radio observations, or else we will set overly stringent constraints based on the smaller angular region considered in the \emph{Galprop} models. We choose the peak flux to stem from the highest flux region with a physical size of at least 25~kpc$^2$, which exceeds the size of the radio beam at our maximum distance for all galaxies. We then round the reported contour down to the next contour level, using the 2$^{n/2}$ contour spacing with the minimum contour flux set by the observation. We note that this assumption is conservative, as the peak radio flux within an observed beam is always slightly higher than the average beam over the matching \emph{Galprop} contour.

In calculations of the minimum flux contour along the minor axis of each galaxy, the anisotropy of the observed emission contours presents a challenge for any determination of the size of the candidate galaxies' minor axis. In order to conservatively estimate the size of the minor axis, we first calculate the direction orthogonal to the observed Galactic plane of each galaxy, and then calculate the smallest minor axis which includes all contours within 10$^\circ$ of the minor axis. We note that in the case of NGC 2683 and NGC 7814, secondary sources are observed in the  -z direction, and thus only the +z measurement is used to set contour constraints. This method effectively chooses the largest possible minor axis which is consistent with the flux data.

\begin{figure}[tbp]
\begin{center}
	\includegraphics[width=\textwidth]{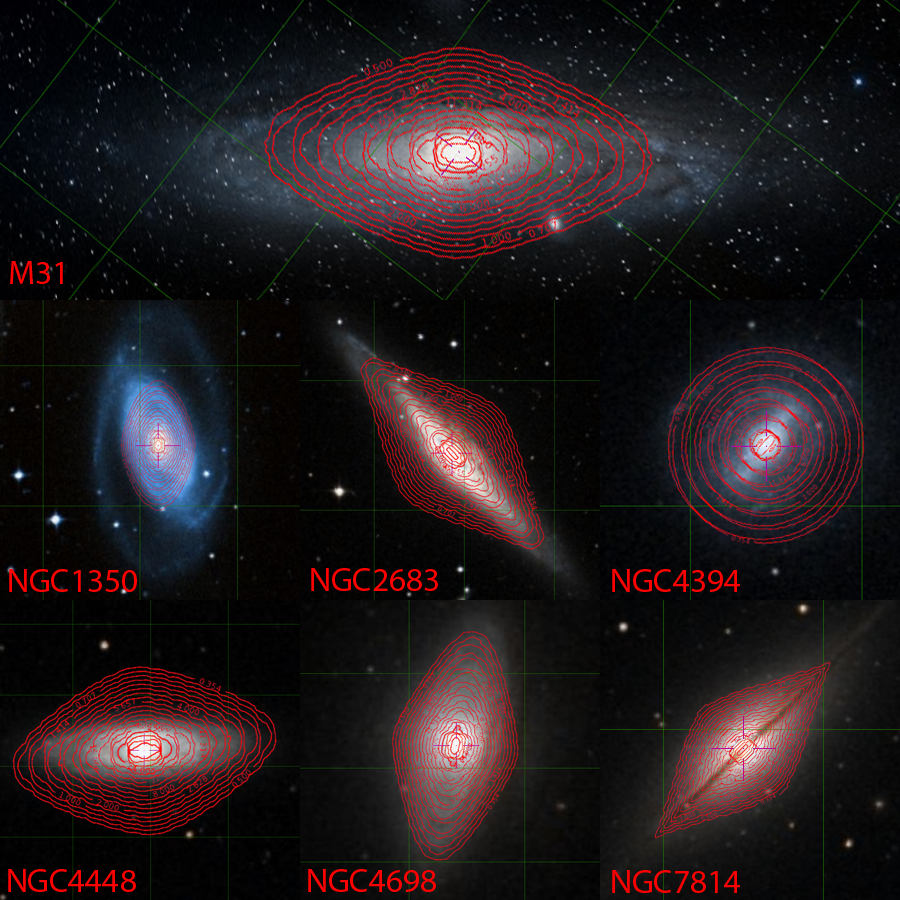}
\end{center}
\caption{Overlays of dark matter only contours from {\em Galprop} onto optical images obtained from the STScI Digitized Sky Survey.  Units are mJy beam$^{-1}$.  Contours are logarithmically spaced at levels $2^{n/2}$ for integer values $n$ with outermost contours $n=-2$ for M31 and $n=-3$ for all others.  With the exception of NGC 4448, which has very large distance uncertainties, the optical dimensions are similar or slightly exceed those of the lowest detectable radio contours.  The diffusion zone has been matched to the estimated physical axis size.}
\label{fig:opticalOverlay}
\end{figure}

In order to verify that our dark matter models are being correctly overlaid onto the radio halo, we also overlay the predicted dark matter contours onto optical images obtained by the STScI Digitized Sky Survey in Figure~\ref{fig:opticalOverlay}. In this case, we consider contours only for the central values of the distance to each candidate galaxy. We observe a strong morphological similarity, which serves to confirm our choices for both the size and orientation of the diffusion region.

\subsection{Constraints from the Integrated Flux}

In Table~\ref{tab:integrated_flux} we show the distance, integrated flux, and the estimate for the dark matter-induced radio haze flux as calculated using \emph{Galprop} for our central parameter values. Also given is the ratio between the observed and simulated flux between 1.49-15~Ghz (R$_{\rm int}$) for each candidate galaxy. Ratios smaller than one, which we highlight in red, indicate that the observed flux is smaller than what we predict {\em from dark matter annihilation alone}. For galaxies with non-negligible distance uncertainties, the uncertainty in the distance, \emph{Galprop} flux, and R$_{\rm int}$ are also given. 

We note two important results: First, R$_{\rm int}$ falls below unity for at least 15 of the 18 galaxy observations, implying that the expected dark matter flux overproduces the entire observed radio flux from each galaxy, which presumably includes both dark matter and astrophysical sources of radio emission. While a constraint on dark matter annihilation from this simple process is difficult to quantitatively determine, we note that from Figure~\ref{fig:scanHist} it is clear that astrophysical cosmic rays usually dominate over the dark matter contribution, implying that for these \EC{7 candidate galaxies}, drawn from a larger distribution of 66 Milky-Way like galaxies, the cosmic ray contribution is extremely low.

Finally, we find that for five galaxies in our sample, $R_{\rm int}$ at 1.49 GHz is smaller than 0.05. In our simulations of the dark matter haze, we find that obtaining a dark matter haze a factor 20 smaller than the benchmark values (i.e. $R_{\rm int}$ smaller than 0.05) has a probability of 11.2\%. The likelihood of having at least 5 galaxies out of a sample of 66 with an emission smaller than 0.05 corresponds to 87.9\%, so we cannot derive any conclusion on the dark matter origin of the haze out of the integrated flux, barring stronger assumptions about the cosmic ray component of the radio emission.  \EC{For clarity we note that by choosing only 7 samples to analyze in detail, our statistical conclusions, which are always taken from the full sample of 66 galaxies, will always be more conservative than performing an in-depth analysis of additional galaxies.}

\begin{table}[tbp]

  \scriptsize
  \centering
  \begin{tabular}{lllllll}
    \toprule
   \multicolumn{1}{c}{Name} & \multicolumn{1}{c}{Distance} & \multicolumn{1}{c}{$\nu$}  & \multicolumn{1}{c}{$\Phi_{\rm Obs}$} & \multicolumn{1}{c}{$\Phi_{\rm GALPROP}$} & \multicolumn{1}{c}{$R_{\rm int}$} & \multicolumn{1}{c}{Ref.} \\
   &\multicolumn{1}{c}{[Mpc]} & \multicolumn{1}{c}{[GHz]}&\multicolumn{1}{c}{[mJy]}&\multicolumn{1}{c}{[mJy]} && \\\midrule \midrule
M31 	& 0.70 			    & 1.49  & 8400     	 & 31967 		      & \textcolor{red}{.263} & \cite{Condon:1987}\\
	 	&     			    & 4.85 	& 1863       & 7840 		      & \textcolor{red}{.238} & \cite{Stil:2009}\\\midrule
NGC1350	& 20.9 			    & 1.49 	& 1.10       & 36.2 		      & \textcolor{red}{.030} & \cite{Condon:1987}\\\midrule
NGC2683 & 10.2 (7.96, 12.4) & 1.49  & 65.9      & 139 (223, 94.0)  & \textcolor{red}{.472} (\textcolor{red}{.295}, \textcolor{red}{.694}) & \cite{Condon:1987}\\
	    & 				    & 2.38  & 39$\pm3$  & 82.6 (137, 55.9)  & \textcolor{red}{.544} (\textcolor{red}{.335}, \textcolor{red}{.805}) & \cite{Condon:1978}\\
	    &	 		 	    & 4.85  & 27$\pm6$  & 34.4 (56.6, 23.3)  & 1.13 (\textcolor{red}{.697}, 1.67) & \cite{Condon:1991}\\
	    &  	                & 15.0  & $<0.9$    & 22.6 (37.1, 15.3)  & \textcolor{red}{.040} (\textcolor{red}{.025}, \textcolor{red}{.059}) & \cite{Nagar:2005}\\\midrule
NGC4394	& 16.8 	            & 1.49  & .7        & 24.1  & \textcolor{red}{.029} & \cite{Condon:1987}\\
		& 	 	            & 2.38  & -1$\pm 4$  & 14.4    &  \textcolor{red}{.486} & \cite{Condon:1978}\\
		& 	 	            & 15.0  & $<.9$  	 & 4.01 & \textcolor{red}{.224} & \cite{Nagar:2005}\\\midrule
NGC4448 & 13.0 (9.70, 47.4) & 1.49  & 1  	     & 70.0 (117, 5.46)  & \textcolor{red}{.014} (\textcolor{red}{.009, .183}) & \cite{Condon:1987}\\
	    & 				    & 2.38  & 0$\pm 4$   & 43.5 (78.1, 3.27)  &  \textcolor{red}{.184} (\textcolor{red}{.102}, 2.44) & \cite{Condon:1978}\\\midrule
	    
NGC4698 & 23.7 (16.9, 30.4) & 1.49  & .6  	     & 26.2 (51.1, 15.8)  & \textcolor{red}{.023} (\textcolor{red}{.012, .038}) & \cite{Condon:1987}\\
    	& 	 	            & 2.38  & 2$\pm 4$   & 15.3 (30.1, 9.29)  & \textcolor{red}{.654} (\textcolor{red}{.334}, 1.08)                  & \cite{Condon:1978}\\
    	& 	 	            & 15.0  & $<1.0$     & 4.19 (8.23, 2.54)  & \textcolor{red}{.240} (\textcolor{red}{.122, 0.394}) & \cite{Nagar:2005}\\\midrule
NGC7814 & 17.2 & 1.49  & 1.1 	     & 49.5  & \textcolor{red}{.022} & \cite{Condon:1987}\\    	
    	& 	 	            & 2.38  & -4$\pm5$   & 29.0  & \textcolor{red}{.207} & \cite{Condon:1978}\\
		& 	 	            & 15.0  & $<0.9$     & 7.95  & \textcolor{red}{.138} & \cite{Nagar:2005}\\
    	
    \bottomrule
  \end{tabular}
  \caption{Integrated flux limits for selected galaxies at available radio frequencies from $1.49-15$ GHz.  First and second values in parentheses represent limits corresponding to the minimum and maximum distance variations respectively. For measurements with associated flux uncertainties, $R_{\rm int}$ was computed using the 95\% confidence level upper limit.  Observations providing constraints are highlighted in red.}
  \label{tab:integrated_flux}
\end{table}

\newcommand{\brp}{\textcolorblack{black}{)}}
\newcommand{\blp}{\textcolor{black}{(}}

\begin{table}[tbp]
  \scriptsize
  \centering
  \begin{tabular}{llllll}
    \toprule
   
   \multicolumn{1}{c}{Name} & \multicolumn{1}{c}{Distance} &  \multicolumn{1}{c}{$R_{\rm peak}$} & \multicolumn{1}{c}{$R_z$} & \multicolumn{1}{c}{$R_r$} \\
   &\multicolumn{1}{c}{[Mpc]} &&&& \\\midrule
M31 	& 0.70 	 & \textcolor{red}{.125}  & \textcolor{red}{.0085 }		    & \textcolor{red}{.0043} \\
NGC1350 & 20.9   & \textcolor{red}{.006} &  \textcolor{red}{.155}  & \textcolor{red}{.0054} \\ 
NGC2683 & 10.2 (7.96, 12.4) &  \textcolor{red}{.063}  & 8.04 (2.36, $\infty$)& \textcolor{red}{.610} (\textcolor{red}{.202}, 2.05)  \\
NGC4394 & 16.8  & \textcolor{red}{.004}  &    \textcolor{red}{.0148}  & \textcolor{red}{.030} \\
NGC4448 & 13.0 (9.70, 47.4) & \textcolor{red}{.004}  & \textcolor{red}{.043 \blp .0172}, $\infty$)& \textcolor{red}{.0173 \blp .0121}, $\infty$) \\
NGC4698 & 23.7 (16.9, 30.4) & \textcolor{red}{.004}  &\textcolor{red}{.0265 \blp .0162}, \textcolor{red}{.107})  & \textcolor{red}{.0034 \blp .0033, .0034})\\
NGC7814 & 17.2 & \textcolor{red}{.008}  & \textcolor{red}{.708} &  \textcolor{red}{.027}\\
    \bottomrule
  \end{tabular}
  \caption{Contour limits for peak, major ($r$), and minor ($z$) fluxes for selected galaxies at $1.49$ GHz.  The first and second values in parentheses represent limits corresponding to the minimum and maximum distance variations respectively.  Observations providing constraints are highlighted in red.}
  \label{tab:peak_and_axis_fluxes}
\end{table}

\subsection{Constraints from Peak and Axis Fluxes}

\begin{figure}[tbp]
\begin{center}

	\includegraphics[scale = 3]{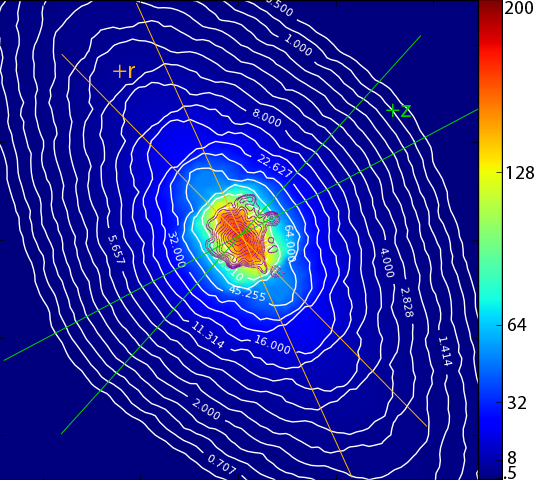}
\end{center}
\caption{10 Kpc square overlay for M31 of Condon Atlas 1.485 GHz radio contours displayed over dark matter only predictions from \emph{Galprop}.  Units are mJy beam$^{-1}$ and the image has been corrected for distance, inclination, position angle, and beam solid angle.  The colormap is linear while contours are logarithmically spaced at levels $2^{n/2}$ for integer values $n$ with outermost contours at the same level $n=-2$.  Contour limits are computed using the conservative case (outermost observed contour) lying within $10^\circ$ major and minor axis (r, z). Although the integrated and peak flux measurements do set limits for M31, the contour limits are the most stringent, constraining a dark matter haze at the $\lesssim 1\%$ level, relative to that of the Milky Way.}
\label{fig:m31overlay}
\end{figure}

While we were not able to obtain any constraints from the total integrated radio luminosity, our dark matter radio emission simulations also allow us to investigate the signal for the galaxies in our sample by examining the peak emission (in the center of the galaxy) and off-peak emission (along and above the galactic plane). In Table~\ref{tab:peak_and_axis_fluxes} we show the flux ratios at 1.49~GHz between each observation and our models at the location of the peak (R$_{\rm peak}$), the minimum observed contour within 10$^\circ$ of the major axis (R$_r$), and the minimum observed contour on within 10$^\circ$ the minor axis (R$_z$). We note four interesting results:
\begin{enumerate}
\item Evaluating the emission at specific locations, instead of the integrated emission, yields stronger constraints in every candidate galaxy.
\item The flux from dark matter annihilation in the majority of our candidate galaxies exceeds the observed radio flux (i.e. $R_{\rm int}<1$) at 1.49~GHz.
\item We find that we get somewhat stronger constraints from R$_r$ than from R$_z$, somewhat contrary to expectations. In the case of the Milky Way the eccentricity in the ellipse stemming from dark matter annihilation is generally smaller than that for modeled astrophysical emission, so one might expect better constraints from off-plane observations.
\end{enumerate}

\begin{figure}[htbp]
\begin{center}
	\includegraphics[width=.9\textwidth]{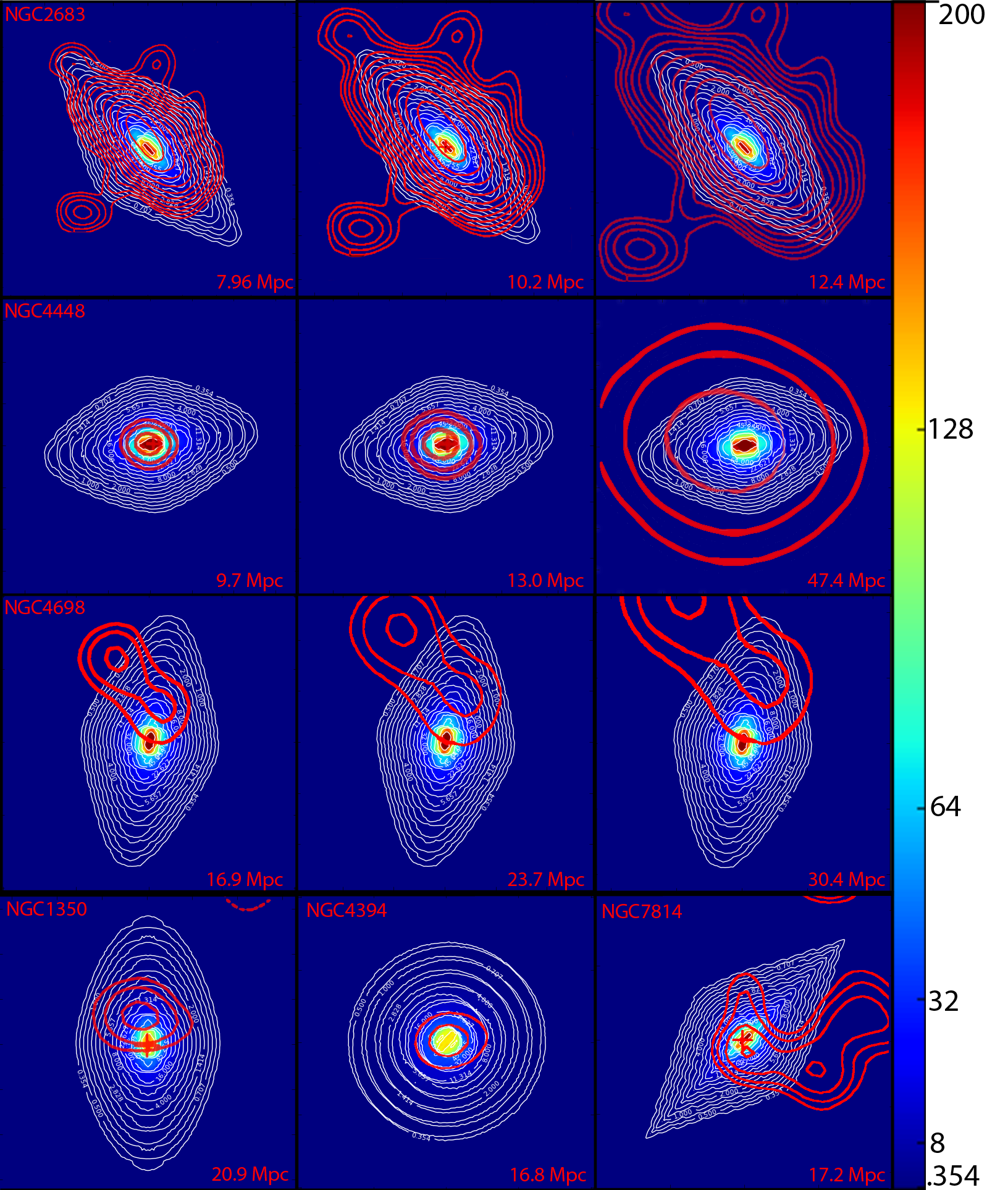}
\end{center}
\caption{20 kpc-square overlays of Condon Atlas 1.485 GHz radio contours onto dark matter only predictions from GALPROP.  Units are mJy beam$^{-1}$ and images have been individually corrected for distance, inclination, position angle, and beam solid angle.  For NGC 4448, distance variations correspond to minimum and maximum NED measurements centered on the value from the Condon Atlas.  For NGC 2683 and 4698, distances are allowed to vary by 1 standard deviation from the mean based on NED measurements.  The colormap is linear while contours are logarithmically spaced at levels $2^{n/2}$ for integer values $n$ with outermost contours at the same level $n=-3$. A dark matter haze should produce a fairly symmetric synchrotron signal.  Here the irregular shapes of the observed contours often yield significant constraints along $r$, $z$, or both.  Additionally, the peak predicted contours are often substantially higher than what is observed.}
\label{fig:overlay}
\end{figure}

In Figures~\ref{fig:m31overlay} and~\ref{fig:overlay} we show the observed 1.485~GHz radio overlays for M31 and for the remainder of our galaxy sample respectively, noting that the flux contours for each follow the form 2$^{n/2}$ mJy~beam$^{-1}$ with an outermost contour of n~=~-2 for M31, and n~=~-3 for the remainder of our candidate galaxies. In Figure~\ref{fig:overlay} we show three distance variations corresponding to the minimum and maximum NED distance measurements for NGC 4448 (2nd row) and the distance estimates corresponding to the 1$\sigma$ distance error for NGC 2683 (top) and NGC 4698 (3rd row). The other three galaxies show a variation of less than 25\% around the mean NED distance, and the variation is not considered.

We note that in many cases the observed flux-contours are highly anisotropic, which contrasts with the approximately symmetric contours expected from dark matter. While anisotropies could be introduced by the cosmic ray propagation parameters, the one-dimensional simulations from Figure~\ref{fig:param} show that these anisotropies in the magnetic field parameters and diffusion constants have relatively minor effects on the radio luminosity, compared to the case of astrophysical injection. Instead, the odd shapes of several galaxies' contours, particularly those of NGC 4698 and NGC 7814, are highly indicative of anisotropic astrophysical cosmic ray injection. Most importantly, we do not see elliptical emission profiles for any galaxy, contrasting with our model for the Milky Way, which has a significant injection of cosmic rays in star formation regions across the Galactic plane. This means that in these radio-underluminous candidate galaxies, the elliptical extension of the expected dark matter profile in the radial direction produces significantly better limits at R$_r$ than in the direction of the minor axis.

The results of Table \ref{tab:peak_and_axis_fluxes} indicate that for three galaxies the off-peak emission along the galaxies' major axis ($r$ direction) is suppressed by more than a factor 100 with respect to our benchmark dark matter haze model. Our simulations, reported as the red line in Figure~\ref{fig:scanHistContour} and in Table~\ref{tab:scan_fractions}, indicate that only 2.3\% of the simulated galaxies are suppressed by at least .006 compared with the benchmark dark matter model.  We calculate that having at least 3 galaxies out of the 66 candidates with such a suppressed radio emission has a likelihood of less than 20\%. Although it is difficult to turn this estimate into a confidence level exclusion limit to the dark matter haze hypothesis, the large variations we allow in multiple physical parameters make it clear that there is some tension in the dark matter interpretation of the galactic haze with observations of external spiral galaxies.

\begin{table}[t]
  \scriptsize
  \centering
  \begin{tabular}{cccccccc}
    \toprule
   Luminosity Cut   & $f_{\rm int}^{1.49\, \textsf{GHz}}$ & $f_{\rm int}^{30 \, \textsf{GHz}}$ & $f_{\rm int}^{44 \,\textsf{GHz}}$ &  $f_{r=5 \,\textsf{kpc}}^{1.49 \,\textsf{GHz}}$ & $f_{z=1 \,\textsf{kpc}}^{1.49\,\textsf{GHz}}$ \\\midrule
    
  $L\leq 0.05 L_{MW,DM}$  & .112  & .249 &  .097  & .152 & .123\\ 
  $L\leq 0.01 L_{MW,DM}$  & .014  & .065 &  .014  & .038 & .017\\
  $L\leq 0.006 L_{MW,DM}$  &  .006 & .034 &  .007  & .023 & .007 \\
    \bottomrule
  \end{tabular}
  \caption{Fractions of random scans with luminosity suppressed by more than a factor of 20 and 100 for a randomized sample of parameter space containing 2000 galaxies with masses restricted to $.25\leq M/M_{MW} \leq 2$.  Subscript ``int'' indicates the total integrated dark matter luminosity and subscript $r$ ($z$) indicates the luminosity evaluated at a given distance along (above) the Galactic plane.}
  \label{tab:scan_fractions}
\end{table}

\subsection{Future directions}

With upcoming data from the \textit{Planck} satellite, we expect to obtain a much more robust analysis of the Milky Way's haze.  \textit{Planck} will produce all-sky observations over nine frequency bands from 30 to 857 GHz, providing a much needed spectrum with superior foreground identification and separation, as well as detailed polarization data.  Having established an effective lower bound on the integrated dark matter luminosity, the availability of an all-sky \emph{spectrum} up to higher frequencies will allow us to transcend the issues associated with complicated cosmic ray sources.  With an angular resolution of 30 arcminutes at 30 GHz and 24 arcminutes at 44 GHz, \textit{Planck} will have the ability to resolve galaxies up to roughly 5 Mpc.  This encompasses the local group which contains several well studied Milky Way analogs.  Galaxies at larger distances will still register as point sources.  If we exclude sources with known nearby radio confusion and cross-reference against a larger galactic catalog to make morphological cuts, it should be possible to obtain integrated flux limits with substantially better statistics than those presented here.

\begin{figure}[tbp]
\begin{center}
	\includegraphics[width=\textwidth]{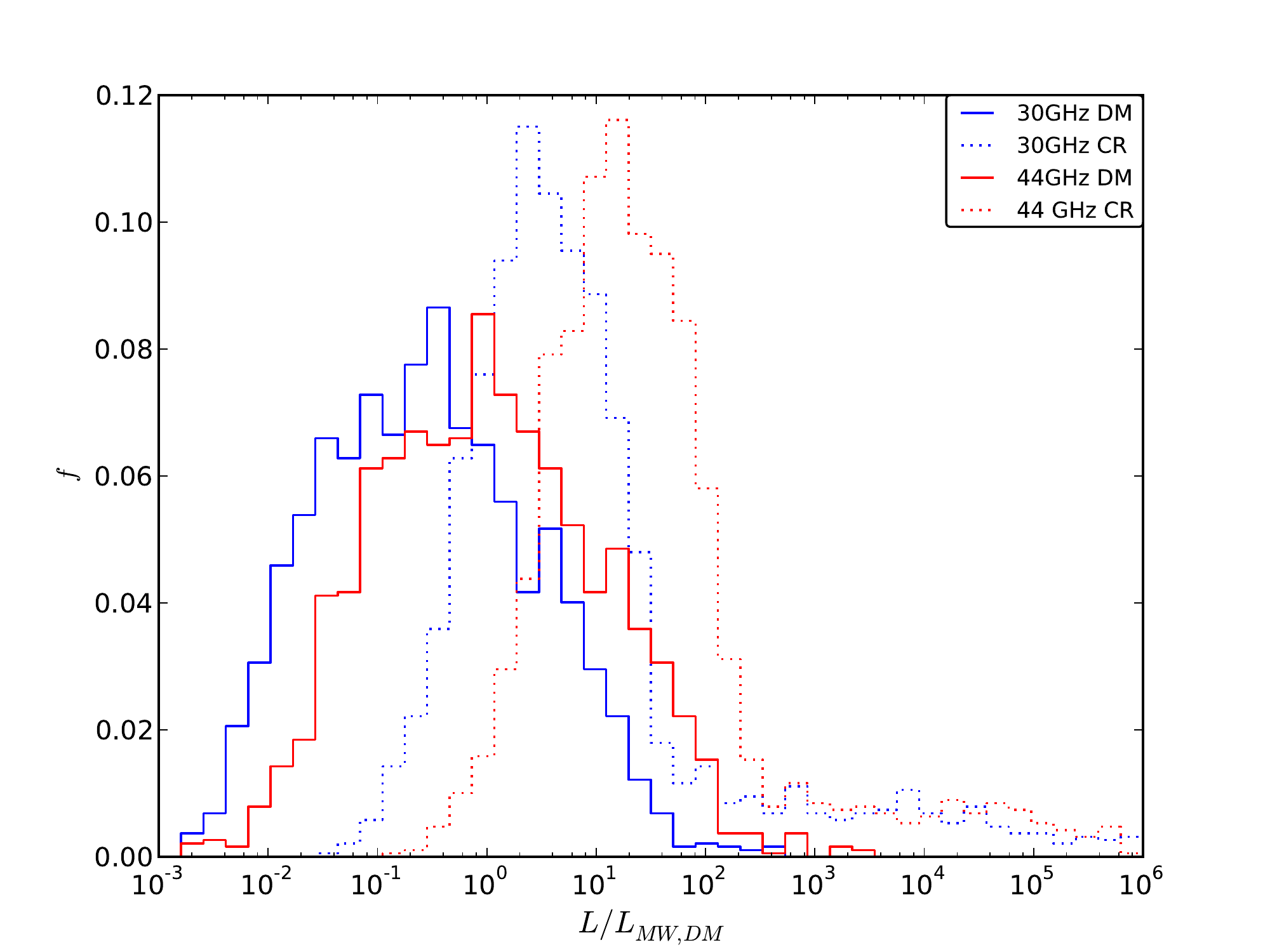}
\end{center}
\caption{Distribution of 30 and 44 GHz luminosity ratios $L/L_{MW,DM}$ where $L$ is the total integrated luminosity and $L_{MW,DM}$ is the luminosity due to dark matter for the canonical Milky Way model at the respective frequency.  Sample contains 2000 runs randomly distributed in the parameter space of Table~\ref{tab:paramrange} with the mass restricted to lie within 0.25 to 2 times that of the Milky Way. We plot the contributions due to dark matter (solid) and cosmic rays (dashed) individually.  At both 30 and 44 GHz, 6.5\% and 1.4\% of models show suppression greater than a factor of 100 compared to the benchmark Milky Way model respectively.  This indicates that higher frequency observations from Planck will produce more robust limits on dark matter origin of the haze.}
\label{fig:scanHistPlanck}
\end{figure}

In order to test the impact of \emph{Planck} observations, Figure \ref{fig:scanHistPlanck} shows the predicted total luminosity components at 30 and 44 GHz, compared to the benchmark Milky Way value for galaxies in our random parameter scan. The distribution is similar in width to that of the 1.49 GHz distribution shown in Figure \ref{fig:scanHist}.  However, the synchrotron spectrum from dark matter is expected to be significantly harder than that of cosmic rays.  Thus, at higher frequencies, we expect a better dark matter signal-to-cosmic ray noise ratio.  Additionally, at 30 GHz (44 GHz), 6.49\% (1.4\%) of models produce a haze with more than a factor 100 suppression compared to the Milky Way model.  With a larger sample, higher quality data, and higher-frequency observations, it should be possible to definitively rule out dark matter annihilation as the primary progenitor of the haze, even under very conservative assumptions.

\section{Discussion and Conclusions}
\label{sec:conclusions}

If the WMAP haze is indeed produced by synchrotron emission from electrons and positrons generated via dark matter annihilation, galaxies with properties similar to those of the Milky Way should also possess an observable radio haze.  High-resolution radio data allows for morphological studies in addition to peak and integrated flux comparisons against simulations.  In this \textit{paper}, we have examined seven galaxies of size and morphological type similar to the Milky Way, and compared the available data at 1.49-15 GHz against the predicted dark matter haze, as calculated using the code {\em Galprop}. Our results are conservative in the sense that we did not subtract the cosmic ray-induced component of the radio emission, which is typically dominant, when setting limits on the dark matter synchrotron emission from the galaxies under consideration.
  
Although there are significant systematic uncertainties involved in this procedure, our analysis found that most of the galaxies under consideration are less luminous at radio frequencies than would be expected from their dark matter-induced haze, unless somewhat extreme parameters are adopted for their dark matter halo profiles, cosmic ray propagation models, or magnetic field models. In particular, to reduce the dark matter luminosity below the observed upper limits, one must simultaneously reduce the dark matter density, increase the diffusion constant, and/or increase the interstellar radiation energy density relative to those features of the Milky Way. The size of the diffusion region and the strength of the magnetic field are less important to this problem.  While most of these parameters are very poorly constrained, we have performed a Monte Carlo scan over what we consider to be a reasonable range of the parameter space, and find that the dark matter-induced haze is less than 1\% as luminous as the Milky Way's haze in only about $\sim5\%$ of the simulated galaxies. 

To obtain limits on the dark matter origin of the haze, we compare observed and predicted flux densities along the semi-major and semi-minor axes, peak contour levels, and integrated flux.  For six of the seven galaxies examined, a dark matter interpretation is constrained below the 5\% level. We obtain stronger constraints from observations at a position of r=5~kpc along the galactic plane, where we find three galaxies which are underluminous by a factor of 100 compared to the Milky Way. Qualitatively, it is also worth noting that the radio emission from the majority of these galaxies does not closely resemble the morphology expected from dark matter annihilation, and instead appears to trace spikes in the cosmic-ray injection rate. Remarkably, one of the galaxies without evidence of a dark matter haze is our local-group neighbor, M31 (the Andromeda Galaxy). While it is difficult to turn this into a firm exclusion limit on the dark matter origin of the Milky Way haze, one would have expected Andromeda to be significantly more radio luminous in most dark matter interpretations of the WMAP haze. 

An important caveat of the analysis in this paper lies in the lack of experimental data on the WMAP haze below 22 GHz.  It is safe to argue that for many dark matter models the approximation of extrapolating linearly from the WMAP frequencies down to 1.49 GHz is valid. In the monochromatic approximation, the synchrotron emission peaks at \cite{Profumo:2013yn}
$$
\frac{\nu_{\rm sync}}{\rm MHz}\simeq2\cdot\left(\frac{E_e}{\rm GeV}\right)\left(\frac{B}{\mu{\rm G}}\right)^{1/2}.
$$
The characteristic electron energy $E_e$ for a dark matter particle is approximately $E_e\sim m_{\rm DM}/10$, so that for typical magnetic fields we expect the peak frequency (where the radio emission is not well-approximated by a power-law) to lie at frequencies between 2 MHz ($m_{\rm DM}\sim$10 GeV) up to 200 MHz ($m_{\rm DM}\sim$1 TeV). Only for masses as large as about 10 TeV do we expect a non-trivial spectral shape for the radio emission between 1.4 GHz and the WMAP frequency range.

In addition, it is worth noting that the magnetic fields within the haze region are highly uncertain, and thus models with very different WIMP masses, may have an equivalent synchrotron spectrum for a different magnetic field structure. In the case of fits to the WMAP haze, this is often done with the intention of fitting the spectrum for WMAP emission between the 22, 33 and 41 GHz frequency bands (see e.g. \cite{lindenhooper}). In this case, the ratio of synchrotron emission at 1.4 and 22 GHz is set entirely by the fit to the WMAP data -- since the Bessel function determining the synchrotron spectrum has only one degree of freedom, which must already be constrained by WMAP measurements. While this statement is necessarily true for various WIMP models annihilating through the same pathway, some minor variations are expected for models which annihilate to softer or harder final states. However, this yields only a very minor adjustment to the relative synchrotron rates, and is not considered in this work.

Upcoming data from the \textit{Planck} space observatory will offer a much broader set of radio frequency observations (nine bands from 30 to 857 GHz). Notably, recent work from the \textit{Planck} collaboration has already confirmed presence of the WMAP haze in the Milky Way, and found that it has a microwave edge spatially coincident with the gamma-ray bubbles detected by \textit{Fermi}. Such a broad frequency range, and additional polarization information, will allow for much greater separation of astrophysical foregrounds, as well as reduced contamination from cosmic rays.

\acknowledgments
We are grateful to A.~Romanowsky for helpful comments and discussions. This work is partly supported by NASA grant NNX11AQ10G. SP acknowledges support from an Outstanding Junior Investigator Award from the Department of Energy, and from DoE grant DE-FG02-04ER41286. 

The Digitized Sky Survey was produced at the Space Telescope Science Institute under U.S. Government grant NAG W-2166. The images of these surveys are based on photographic data obtained using the Oschin Schmidt Telescope on Palomar Mountain and the UK Schmidt Telescope. The plates were processed into the present compressed digital form with the permission of these institutions.

This research has made use of the NASA/IPAC Extragalactic Database (NED) which is operated by the Jet Propulsion Laboratory, California Institute of Technology, under contract with the National Aeronautics and Space Administration.

\begin{landscape}
{\footnotesize
\begin{longtable}[htbp]{ccccccccccccc}

\caption{Data on Selected Galaxies}\\

\hline
               Name                   &       Type        &         t         &       i        &       $v_{rot}$        &  \multicolumn{1}{c}{D} &\multicolumn{1}{c}{S} & \multicolumn{1}{c}{Sp} & \multicolumn{1}{c}{S$_p$*D$^2$ (2kpc)}  & \multicolumn{1}{c}{S D$^2$} & \multicolumn{1}{c}{Rank L$_I$} & \multicolumn{1}{c}{Rank 2kpc} & \multicolumn{1}{c}{Mean Rank} \\                
 &  &  & [$^\circ$] & [km/s] & \multicolumn{1}{c}{[Mpc]} & \multicolumn{1}{c}{[mJy]} & \multicolumn{1}{c}{[mJy]} & \multicolumn{1}{c}{[mJy Mpc]}  & \multicolumn{1}{c}{[mJy Mpc]} & \multicolumn{1}{c}{} & \multicolumn{1}{c}{} & 
\multicolumn{1}{c}{} \\

\hline
\endfirsthead
\multicolumn{13}{c}%
{\tablename\ \thetable\ -- \textit{Continued from previous page}} \\
\hline
               Name                   &       Type        &         t         &       i        &       $v_{rot}$        &  \multicolumn{1}{c}{D} &\multicolumn{1}{c}{S} & \multicolumn{1}{c}{Sp} & \multicolumn{1}{c}{S$_p$*D$^2$ (2kpc)}  & \multicolumn{1}{c}{S*D$^2$} & \multicolumn{1}{c}{Rank L$_I$} & \multicolumn{1}{c}{Rank 2kpc} & \multicolumn{1}{c}{Mean Rank} \\                
 &  &  & [$^\circ$] & [km/s] & \multicolumn{1}{c}{[Mpc]} & \multicolumn{1}{c}{[mJy]} & \multicolumn{1}{c}{[mJy]} & \multicolumn{1}{c}{[mJy Mpc$^2$]}  & \multicolumn{1}{c}{[mJy Mpc$^2$]} & \multicolumn{1}{c}{} & \multicolumn{1}{c}{} & 
\multicolumn{1}{c}{} \\
\hline
\endhead
\hline \multicolumn{13}{r}{\textit{Continued on next page}} \\
\endfoot
\hline
\endlastfoot


NGC 4448                                 &  SBab             &  1.8              &  52.51            &  221.54           & 13.1 & 1 & 0.9  & 2.42E-6 & 2.16E+3 & 1 & 2 & 1.5 \\ 
NGC 4698                                 &  Sab              &  1.7              &  73.44            &  201.13           & 21.9 & 0.6 & \multicolumn{1}{l}{      } & 0.00E+0 & 6.03E+3 & 2 & \multicolumn{1}{l}{} & 2 \\ 
NGC 4394                                 &  SBb              &  3                &  16.55            &  255.13           & 21.9 & 0.7 & 0.6 &   4.82E-6 & 6.03E+3 & 3 & 3 & 3 \\ 
NGC 7814                                 &  Sab              &  2                &  70.59            &  230.9            & 25.8 & 1.1 & 1.1 &   9.63E-6 & 8.36E+3 & 4 & 6 & 5 \\ 
NGC 1350                                 &  Sab              &  1.9              &  64.79            &  199.87           & 30.2 & 1.1 & 0.8 &   9.63E-6 & 1.15E+4 & 5 & 7 & 6 \\ 
NGC 0224                                 &  Sb               &  3                &  72.17            &  256.7            & 0.7 & 8400 & 14.5 &   8.99E-8 & 6.16E+0 & 12 & 1 & 6.5 \\ 
NGC 2683                                 &  Sb               &  3                &  82.79            &  202.92           & 8 & 65.9 & 14.6 &   9.12E-6 & 8.04E+2 & 13 & 5 & 9 \\ 
NGC 3031                                 &  Sab              &  2.4              &  62.69            &  216.59           & 3.6 & 380 & 85 &   2.21E-5 & 1.63E+2 & 14 & 8 & 11 \\ 
NGC 4274                                 &  SBab             &  1.7              &  68.01            &  236.74           & 17.6 & 11.2 & 9.5 &   5.90E-5 & 3.89E+3 & 8 & 18 & 13 \\ 
NGC 4216                                 &  SABb             &  3                &    90             &  243.98           & 21.9 & 13.4 & 4.3 &   3.34E-5 & 6.03E+3 & 18 & 9 & 13.5 \\ 
NGC 3898                                 &  Sab              &  1.7              &  56.26            &  263.55           & 25.2 & 6.3 & 4.3 &   5.60E-5 & 7.98E+3 & 11 & 17 & 14 \\ 
NGC 3368                                 &  SABa             &  2.2              &  51.04            &  203.88           & 15.2 & 30 & 10.5 &   4.07E-5 & 2.90E+3 & 19 & 11 & 15 \\ 
NGC 4450                                 &  Sab              &  2.4              &  48.75            &  193.62           & 21.9 & 8.3 & 7.5 &   6.03E-5 & 6.03E+3 & 10 & 20 & 15 \\ 
NGC 4013                                 &  Sb               &  3.1              &    90             &  181.68           & 17.3 & 34.6 & 13.9   & 4.35E-5 & 3.76E+3 & 24 & 13 & 18.5 \\ 
NGC 3675                                 &  Sb               &  3                &  59.45            &  222.59           & 15.8 & 43.7 & 17.2  & 5.52E-5 & 3.14E+3 & 25 & 15 & 20 \\ 
NGC 3344                                 &  Sbc              &  4                &  18.72            &  222.87           & 12.5 & 85.5 & 11.5  & 3.73E-5 & 1.96E+3 & 30 & 10 & 20 \\ 
NGC 4736                                 &  Sab              &  2.4              &  31.74            &  182.09           & 6.9 & 254 & 61.4 &   4.85E-5 & 5.98E+2 & 28 & 14 & 21 \\ 
NGC 5566                                 &  SBab             &  1.6              &  75.57            &  202.28           & 29.4 & 8.5 & 4.6 &   6.64E-5 & 1.09E+4 & 20 & 23 & 21.5 \\ 
NGC 4941                                 &  SABa             &  2.1              &  36.6             &  229.7            & 17.6 & 19 & 18.8 &   1.12E-4 & 3.89E+3 & 16 & 28 & 22 \\ 
NGC 4051                                 &  SABb             &  4                &  30.22            &  213.13           & 14.9 & 77 & 11.7 &   4.29E-5 & 2.79E+3 & 34 & 12 & 23 \\ 
NGC 7217                                 &  Sab              &  2.5              &  33.42            &  258.91           & 24.7 & 16.2 & 10.5 &   8.49E-5 & 7.67E+3 & 23 & 24 & 23.5 \\ 
NGC 3953                                 &  Sbc              &  4                &  62.14            &  215.86           & 20.7 & 41.1 & 7.2 &  6.37E-5 & 5.38E+3 & 35 & 21 & 28 \\ 
NGC 4699                                 &  SABb             &  2.9              &  42.61            &  258.46           & 26.2 & 15.9 & 8.5 &   1.20E-4 & 8.63E+3 & 26 & 31 & 28.5 \\ 
NGC 4725                                 &  SABa             &  2.2              &  45.39            &  257.47           & 23.3 & 28.2 & \multicolumn{1}{l}{      } &   0.00E+0 & 6.82E+3 & 31 & \multicolumn{1}{l}{} & 31 \\ 
NGC 7723                                 &  SBb              &  3.1              &  47.09            &  192.3            & 39.5 & 10.4 & 7.2   & 1.50E-4 & 1.96E+4 & 33 & 34 & 33.5 \\ 
IC 0520                                  &  SABa             &  1.7              &  41.38            &  211.51           & 73.2 & 3 & 2.5 &   1.80E-4 & 6.73E+4 & 32 & 39 & 35.5 \\ 
NGC 2903                                 &  SABb             &  4                &  67.09            &  186.95           & 9.4 & 407 & 73.7   & 1.33E-4 & 1.11E+3 & 41 & 32 & 36.5 \\ 
NGC 5055                                 &  Sbc              &  4                &  54.87            &  218.43           & 11 & 390 & 56.9 &   1.16E-4 & 1.52E+3 & 45 & 30 & 37.5 \\ 
NGC 4192                                 &  SABb             &  2.6              &    90             &  230.81           & 21.9 & 73.9 & 22.4   & 1.79E-4 & 6.03E+3 & 40 & 38 & 39 \\ 
NGC 1532                                 &  SBb              &  3.1              &  84.17            &  257.85           & 22.1 & 95.8 & 17   & 1.62E-4 & 6.14E+3 & 44 & 35 & 39.5 \\ 
NGC 2336                                 &  Sbc              &  4                &  58.31            &  242.67           & 48.5 & 17.7 & \multicolumn{1}{l}{      }   & 0.00E+0 & 2.96E+4 & 42 & \multicolumn{1}{l}{} & 42 \\ 
NGC 2935                                 &  Sb               &  3.1              &  42.94            &  187.41           & 40.1 & 40 & 6.8 &   1.47E-4 & 2.02E+4 & 54 & 33 & 43.5 \\ 
NGC 6384                                 &  SABb             &  3.6              &  66.56            &  182.54           & 34.7 & 34.9 & 6.8  & 2.48E-4 & 1.51E+4 & 43 & 44 & 43.5 \\ 
NGC 3628                                 &  Sb               &  3.1              &  79.3             &  215.34           & 14.4 & 525 & 33.6   & 1.14E-4 & 2.61E+3 & 64 & 29 & 46.5 \\ 
NGC 4157                                 &  SABb             &  3.3              &    90             &  188.89           & 17 & 180 & 55.8   & 2.36E-4 & 3.63E+3 & 50 & 43 & 46.5 \\ 
NGC 3521                                 &  SABb             &  4                &  60.02            &  244.92           & 12.5 & 357 & 71.8   & 2.29E-4 & 1.96E+3 & 52 & 42 & 47 \\ 
NGC 5792                                 &  Sb               &  3                &  72.4             &  210.81           & 37.8 & 51.8 & 8   & 1.92E-4 & 1.80E+4 & 56 & 40 & 48 \\ 
NGC 0278                                 &  SABb             &  3                &  12.76            &  256.28           & 18.6 & 138 & 71   & 3.05E-4 & 4.35E+3 & 47 & 49 & 48 \\ 
NGC 4565                                 &  Sb               &  3.3              &    90             &  244.91           & 23.4 & 131 & 18   & 2.02E-4 & 6.88E+3 & 55 & 41 & 48 \\ 
NGC 3982                                 &  SABb             &  3.2              &  29.9             &  191.83           & 23.9 & 61 & 44   & 5.07E-4 & 7.18E+3 & 39 & 59 & 49 \\ 
NGC 1964                                 &  SABb             &  3.3              &  67.08            &  206.36           & 31.6 & 47.8 & 31   & 4.10E-4 & 1.25E+4 & 46 & 53 & 49.5 \\ 
NGC 4217                                 &  Sb               &  3.1              &  81.01            &  187.64           & 21.7 & 109 & 43.9   & 3.25E-4 & 5.92E+3 & 49 & 52 & 50.5 \\ 
NGC 0289                                 &  SBbc             &  4                &  42.95            &  184.3            & 36.7 & 47 & 17.6  & 3.17E-4 & 1.69E+4 & 53 & 51 & 52 \\ 
NGC 5005                                 &  SABb             &  4                &  77.05            &  250.19           & 20.8 & 176 & 49.5   & 2.78E-4 & 5.44E+3 & 57 & 48 & 52.5 \\ 
NGC 4579                                 &  SABb             &  2.9              &  41.86            &  248.63           & 21.9 & 103 & 62.9   & 5.10E-4 & 6.03E+3 & 48 & 60 & 54 \\ 
NGC 2985                                 &  Sab              &  2.4              &  37.88            &  224.73           & 29 & 61.9 & 28.1 &  4.93E-4 & 1.06E+4 & 51 & 58 & 54.5 \\ 
NGC 0891                                 &  Sb               &  3.1              &    90             &  212.09           & 15.6 & 701 & 84.8 &  2.61E-4 & 3.06E+3 & 68 & 45 & 56.5 \\ 
NGC 1365                                 &  Sb               &  3.2              &  62.71            &  198.16           & 30.2 & 530 & 13.3 &   1.63E-4 & 1.15E+4 & 77 & 36 & 56.5 \\ 
NGC 4939                                 &  Sbc              &  4                &  70.07            &  220.3            & 58.1 & 24.2 & 10 &   4.55E-4 & 4.24E+4 & 59 & 55 & 57 \\ 
NGC 4258                                 &  SABb             &  4                &  68.33            &  212.96           & 10.4 & 790 & \multicolumn{1}{l}{      }   & 0.00E+0 & 1.36E+3 & 60 & \multicolumn{1}{l}{} & 60 \\ 
NGC 3223                                 &  Sb               &  3.4              &  51.32            &  246.68           & 52.4 & 32.2 & 11.1  & 5.19E-4 & 3.45E+4 & 61 & 61 & 61 \\ 
NGC 4041                                 &  Sbc              &  4                &  22.02            &  263.1            & 27.2 & 103 & 69.7 & 8.62E-4 & 9.30E+3 & 58 & 64 & 61 \\ 
NGC 1097                                 &  SBb              &  3.3              &  54.95            &  219.98           & 25.7 & 415 & 19.9  & 2.77E-4 & 8.30E+3 & 76 & 47 & 61.5 \\ 
NGC 0134                                 &  SABb             &  4                &    90             &  220.22           & 31.9 & 191 & 30.7   & 4.15E-4 & 1.28E+4 & 70 & 54 & 62 \\ 
NGC 5371                                 &  Sbc              &  4                &  53.97            &  222.5            & 52.3 & 50.3 & 10.3   & 4.81E-4 & 3.44E+4 & 67 & 57 & 62 \\ 
NGC 7331                                 &  Sbc              &  3.9              &  69.96            &  252.42           & 22.3 & 373 & 71.8   & 4.76E-4 & 6.25E+3 & 69 & 56 & 62.5 \\ 
NGC 1055                                 &  SBb              &  3.1              &  62.5             &  204.32           & 22 & 213 & 99.7   & 9.91E-4 & 6.08E+3 & 62 & 66 & 64 \\ 
NGC 4030                                 &  Sbc              &  4                &  47.07            &  201.32           & 26.4 & 151 & 66.5   & 9.63E-4 & 8.76E+3 & 63 & 65 & 64 \\ 
NGC 4501                                 &  Sb               &  3.3              &  62.94            &  272.19           & 21.9 & 278 & 73.7   & 5.95E-4 & 6.03E+3 & 66 & 63 & 64.5 \\ 
NGC 4902                                 &  Sb               &  3                &  26.61            &  250.23           & 48.5 & 52 & 24 &  1.19E-3 & 2.96E+4 & 65 & 67 & 66 \\ 
NGC 4303                                 &  Sbc              &  4                &  18.06            &  213.78           & 21.9 & 416 & 67.4   & 5.49E-4 & 6.03E+3 & 71 & 62 & 66.5 \\ 
NGC 0772                                 &  Sb               &  3.1              &  59.59            &  260.73           & 52.9 & 71.4 & 27.2   & 1.61E-3 & 3.52E+4 & 72 & 68 & 70 \\ 
NGC 3504                                 &  Sab              &  2.1              &  26.15            &  194.09           & 29.6 & 265 & 234  & 3.41E-3 & 1.10E+4 & 73 & 70 & 71.5 \\ 
NGC 7582                                 &  SBab             &  2.1              &  68.17            &  194.86           & 29.5 & 270 & 220   & 3.93E-3 & 1.09E+4 & 74 & 71 & 72.5 \\ 
NGC 7552                                 &  Sab              &  2.4              &  23.64            &  205.77           & 31.3 & 276 & 221   & 4.48E-3 & 1.23E+4 & 75 & 72 & 73.5 \\ 
NGC 6907                                 &  Sbc              &  4                &  37.46            &  212.88           & 63.8 & 136 & 60.6   & 3.33E-3 & 5.12E+4 & 78 & 69 & 73.5 \\ \bottomrule


\label{tab:selected}
\end{longtable}
}
\end{landscape}

\bibliography{ms}
\bibliographystyle{JHEP}

\end{document}